%% file: FPGA2021.tex
\documentclass[sigconf]{acmart}

\AtBeginDocument{%
  \providecommand\BibTeX{{%
    \normalfont B\kern-0.5em{\scshape i\kern-0.25em b}\kern-0.8em\TeX}}}

\copyrightyear{2021}
\acmYear{2021}
\setcopyright{usgovmixed}\acmConference[FPGA '21]{Proceedings of the 2021 ACM/SIGDA International Symposium on Field Programmable Gate Arrays}{February 28-March 2, 2021}{Virtual Event, USA}
\acmBooktitle{Proceedings of the 2021 ACM/SIGDA International Symposium on Field Programmable Gate Arrays (FPGA '21), February 28-March 2, 2021, Virtual Event, USA}
\acmPrice{15.00}
\acmDOI{10.1145/3431920.3439286}
\acmISBN{978-1-4503-8218-2/21/02}

\usepackage{outlines}

\usepackage{booktabs}
\usepackage{amsfonts,amsmath}
\usepackage{algorithm}% http://ctan.org/pkg/algorithms
\usepackage[noend]{algpseudocode}% http://ctan.org/pkg/algorithmicx
\usepackage{bm}
\usepackage{booktabs}
\usepackage{multirow}
\usepackage{subfig}
\usepackage{soul}
\usepackage{tikz}
\newcommand*\circled[1]{\tikz[baseline=(char.base)]{
            \node[shape=circle,draw,inner sep=1.1pt] (char) {#1};}}
\usepackage[flushleft]{threeparttable}

\definecolor{mygreen}{RGB}{26, 132, 62}
\begin{document}

\title{DYNAMAP: \underline{Dyna}mic Algorithm \underline{Map}ping Framework for \\Low Latency CNN Inference}
\author{Yuan Meng}
\email{ymeng643@usc.edu}
\affiliation{%
  \institution{University of Southern California}
}
\author{Sanmukh Kuppannagari}
\email{kuppann@usc.edu}
\affiliation{%
  \institution{University of Southern California}
}
\author{Rajgopal Kannan}
\email{rajgopal.kannan.civ@mail.mil}
\affiliation{%
  \institution{US Army Research Lab}
}
\author{Viktor Prasanna}
\email{prasanna@usc.edu}
\affiliation{%
 \institution{University of Southern California}}
\begin{abstract}
  Most of the existing work on FPGA acceleration of Convolutional Neural Network (CNN) focus on employing a single strategy (algorithm, dataflow, etc.) across all the layers.
  Such an approach does not achieve optimal latency on complex and deep CNNs. Emerging CNNs have diverse per-layer computation characteristics including parallelism, arithmetic intensity, locality, and memory footprint. Per-layer strategy selection and 
  fine-grained tuning is required to achieve low end-to-end latency.
However, specialized hardware modules dedicated for each layer limit the per-layer utilization and adversely affect end-to-end latency.
  In this paper, we address these problems by an algorithm-architecture co-optimization framework, DYNAMAP, consisting of (1) a unified hardware overlay that can be reused across layers, supporting dynamic mapping of all three families of popular convolution algorithms, and further allowing flexible dataflow switching to maximize hardware utilization for each layer; (2) a novel software Design Space Exploration (DSE) flow that customizes the hardware overlay and chooses optimal strategy mapping.
 We show that the algorithm mapping space increases exponentially with network depth, and while the optimal algorithm selection problem is NP-hard in general, by exploiting the series-parallel structure of CNN models, we demonstrate a polynomial-time solution for optimal algorithm mapping.
DYNAMAP is optimized for any CNN, including those having diverse computation and memory requirements across the layers.
  We demonstrate DYNAMAP using two state-of-the-art CNNs - GoogleNet and Inception-V4. The generated accelerators achieve up to $2.8\times$ and $1.4\times$ speedups, respectively, wrt inference latency compared with the state-of-the-art FPGA implementations.
\end{abstract}
\maketitle
\input{1_intro}
\input{2_background}
\input{3_Architectural_Design}
\input{4_pbqp}

\input{5_DSE}

\input{6_Experiments}
\input{8_conclusion}

\bibliographystyle{ACM-Reference-Format}
\bibliography{FPGA2021}

\end{document}

%% file: 1_intro.tex
\section{Introduction}
CNNs are powerful techniques used in many computer vision related machine learning tasks spanning from image and video processing to automatic learning of agents in robotics and games. 
Recently, new families of CNNs featuring various convolution (CONV) layers  \cite{szegedy2015going,szegedy2017inception,howard2017mobilenets} have been developed. These CNNs show superior performance wrt prediction accuracy while introducing new convolution operations (e.g. depthwise CONV in MobileNet, "Fire" module in SqueezeNet, $1\times 7$ filter in "Inception module"). 
As CNNs continue to be more compute- and memory-intensive,
FPGAs become promising candidates for CNN inference implementation with superior latency and reduced energy consumption. 
To avoid expensive run-time reconfiguration, most existing FPGA-based solutions usually use a specific algorithm across all the layers and reuse a generic architecture to speed up the algorithm \cite{zeng2018framework,podili2017fast,niu2020reuse,zhang2020evaluating, ye2020hybriddnn}. These approaches leave some performance and hardware efficiency on the table due to (1) use of a fixed algorithm across diverse layers and (2) under-utilization due to fixed hardware. 
This situation is compounded since state-of-the-art CNNs have more diverse layer shapes and complex structures, resulting in sub-optimal latency. 

Many algorithms have been proposed to optimize various types of CONV operations, such as GEMM (General Matrix-Matrix Multiplication) based methods (im2col \cite{chetlur2014cudnn}, kn2row \cite{vasudevan2017parallel}, Winograd-fast matrix multiplication \cite{lavin2016fast}) and frequency-domain methods \cite{wang2016cnnpack}.
DYNAMAP is primarily motivated by the possibility of bridging the performance gap between the widely used "one size fits all" design methodology and an ideal design methodology where \textit{dedicated layer-wise algorithm tuning is performed to realize low end-to-end latency
and maximal hardware-reuse across layers}. To achieve this goal, this paper explores dynamic algorithm mapping for various layers using a domain specific architecture that can support all the GEMM-based CONV algorithms.

Different algorithmic approaches and layer configurations introduce different trade-offs wrt parallelism, memory requirements and data layout which can easily cause hardware under-utilization. 
Deeper and increasingly complex CNN models lead to exponential explosion of algorithm design space. For example, GoogleNet has 22 CONV layers. Assuming each layer can be implemented with 3 algorithm choices, this leads to an algorithm mapping space of over 30 billion ($3^{22}$) combinatorial choices. 
 It is non-trivial effort to decide which algorithm mapping is the best for a given CNN model on a given hardware - This assignment problem naturally maps to Partitioned Boolean Quadratic Programming (PBQP) \cite{scholz2002register} optimization problem, which is NP-hard in its general form.
To address the challenge of hardware under-utilization, we apply dataflow optimization
to minimize wasted computation. To achieve fast optimal algorithm mapping, we propose a polynomial-time solution taking advantage of the series-parallel graph structure of CNNs. These are integrated into DYNAMAP, which takes as inputs (a) the CNN model parameters (input and kernel shapes, stride sizes) (b) the set of algorithms,
and (c) FPGA device meta data (DSP resources, on-chip memory size and external bandwidth). The outputs are algorithm selection for each layer and customized accelerator on the target device specified in Verilog.

Our main contributions are:
\begin{outline}
\1 A unified template hardware overlay re-used across layers with the following novel optimizations:
    \2 \textbf{Algorithm Switching}: We enable low-overhead layer-wise layout transformation, allowing dynamic switching of three popular GEMM-based CONV algorithms: im2col, kn2row and Winograd algorithms; 
    \2 \textbf{Dataflow Optimization}: We optimize the Processing Elements (PE) design to support no-overhead switching between different dataflows to achieve maximum hardware utilization, enhanced with conflict-free memory layout;
\1 Accurate modeling of the computation and communication latency for various CONV algorithm combinations
to allow easy construction of a parameterized dependency graph representation for any CNN models, capturing architectural parameters, FPGA device capabilities, and CNN meta data;
\1 A framework, DYNAMAP, that proposes a 2-step DSE flow:
    \2 \textbf{Hardware Mapping}:
    Identifying the fixed architectural parameters as well as the most efficient dataflow for each layer under all algorithm settings;
    \2 \textbf{Algorithm Mapping}: Polynomial-time optimal algorithm selection exploiting the series-parallel characteristic of CNN graphs;
\1 We evaluate the DYNAMAP-generated hardware design on a Xilinx Alveo U200 board. Our designs achieve up to $2.8\times$ and $1.4\times$ end-to-end latency improvements compared with state-of-the-art FPGA implementations on two recent deep complex-form CNNs - GoogleNet and Inception-V4. 
\end{outline}

%% file: 2_background.tex
\section{Background \& Motivation}
\subsection{GEMM-based Convolution}
Convolution (CONV) layers are major building blocks of CNNs and their meta data are defined as follows: Each CONV layer has  $C_{in}$ ($C_{out}$) input (output) channels, where each channel is a $H_1\times H_2$ ($O_1\times O_2$) 2D feature map. The layer weights $W$ contain $C_{in}\times C_{out}$ number of kernels, each sized at $K_1\times K_2$. In this work, we focus on spatial convolution. Spatial convolution performs sliding window multiply-accumulate operation of the kernels over the feature maps. As CONV layers dominate the memory and computations in a CNN, a number of algorithms have been proposed for efficient implementation of convolution operation. Among these General-matrix-multiplication (GEMM) - based methods are most widely adopted \cite{chetlur2014cudnn,vasudevan2017parallel,lavin2016fast} for spatial convolution.  In this section, we summarize three families of popular GEMM-CONV algorithms and their trade-offs.

\subsubsection{im2col Method}
im2col \cite{chetlur2014cudnn} is a popular algorithm that converts spatial convolution into GEMM. For a feed forward pass of a CONV layer:

\begin{equation}
\boldsymbol{z}_{x, y}^{l}=\sum_{x^{\prime}} \sum_{y^{\prime}} \boldsymbol{w}_{x, y}^{l} \boldsymbol{a}_{\left(x+x^{\prime}\right),\left(y+y^{\prime}\right)}^{l-1} 
\end{equation}

im2col stretches each group of $C_{in}$ kernels into a row of the weight/kernel matrix, $\boldsymbol{X}$, and each group of $C_{in}$ corresponding windows of input feature maps into a column of the input activation matrix, $\boldsymbol{W}$, expressing the feed forward pass as

\begin{equation}
\boldsymbol{z}_{l}=\boldsymbol{W}^{(C_{out}\times K_1K_2C_{in})}_l* \boldsymbol{X}^{(K_1K_2C_{in}\times O_1O_2)}_{l-1}
\end{equation}

\subsubsection{kn2row Method}
The kn2row \cite{vasudevan2017parallel} method is based on the decomposition of convolutions and
reordering the data layout. In the first phase, "unit-CONV GEMM", a $K_1\times K_2$ convolution is computed using $K_1K_2$ separate $1\times 1$ unit-convolutions, which is equivalent to a GEMM call:

\begin{equation}
\boldsymbol{p}_{k_1,k_2}=\boldsymbol{W}_l^{(C_{out}\times C_{in})}* \boldsymbol{X}_{l-1}^{(C_{in}\times O_1O_2)}
\end{equation}
In the second phase, "Pad-and-Accumulate", the intermediate output patches of all unit-convolutions, $\boldsymbol{p}_{k_1,k_2}$, are shifted by their offsets w.r.t. the center patch, padded with $0$ on the non-overlapping areas and Hadamard-added to generate the final output feature maps:
\begin{equation}
\vspace{-2mm}
\boldsymbol{z}_{l}^{x,y}=\sum_{k_1=1,k_2=1}^{K_1,K_2}\boldsymbol{p}_{k_1,k_2}^{x+(k_1-\frac{K_1}{2}),y+(k_2-\frac{K_2}{2})}
\end{equation}
\subsubsection{Winograd Minimal Filtering Method}
Winograd algorithm \cite{lavin2016fast} is a fast matrix multiplication algorithm which reduces the number of operations in a GEMM call. A $F(m\times m, r\times r )$ Winograd algorithm can generate output as Equation \ref{winograd}. In this equation, $Y$ and $d$ represent output and input tiles, while $g$ represents kernels. $A, G,$ and $B$ are constant transforming matrices and $\odot$ represents Hadamard multiplication. Input feature map $D$ is partitioned into multiple input tiles, $d$, with size $(m+r-1) \times(m+r-1)$ while adjacent tiles share an $(r-1)$ overlap. Each output tile size is $m \times m$ and kernel size is $r \times r$.
\begin{equation}
\vspace{-2mm}
\label{winograd}
Y=A^{T}\left[\left[G g G^{T}\right] \odot\left[B^{T} d B\right]\right] A
\end{equation}
The final output feature map is calculated by concatenating all output tiles, $Y$, and summing up over the depth of the input tensor. Equivalently, we can reduce over $C_{in}$ channels in the transform space before applying the inverse transform $A$ to the sum. This amortizes the cost of the inverse transform over the number of channels, and allows us to alternatively express the Hadamard products and depth-wise additions into $(m+r-1) \times(m+r-1)$ independent GEMMs \cite{lavin2016fast}:
\vspace{-2mm}
\begin{equation}
\vspace{-2mm}
    M_{k, \widetilde{x}, \widetilde{y}}^{(\xi, \nu)}=\sum_{c=1}^{C_{in}} U_{k, c}^{(\xi, \nu)} V_{c, \widetilde{x}, \widetilde{y}}^{(\xi, \nu)}\\
\end{equation}
where we label each single component of the tiled Hadamard multiplication separately, as $(\xi, \nu)$ ($0<\xi, \nu\leq m+r-1$); $U$ and $V$ represent the input and kernel tiles $G g G^{T}$ and $B^{T} d B$, respectively; and the tuple $k, (\widetilde{x}, \widetilde{y})$ indicates the coordinate of the input tile corresponding to $k^{th}$ kernel ($0<k\leq C_{out}$).

\textbf{Trade-offs:} While im2col has the same computation load as spatial convolution, it suffers from memory overhead due to input feature duplication, especially for large kernels and small stride sizes. kn2row is known as a low-memory algorithm because it eliminates the need to duplicate feature map elements by using $1\times 1$ convolutions. However, it incurs extra overhead in the "Pad-and-Accumulate" phase compared to im2col method. The advantage of Winograd CONV comes from the lower computation complexity. For example, an $F(4 \times 4,3 \times 3)$ Winograd algorithm requires 36 multiplications for one output tile, while the Spatial CONV needs 144 multiplications. The reduction of multiplications is 4 times in this case. However, Winograd introduces memory overhead and extra additions due to the transformation in Equation \ref{winograd}. 

\subsection{Motivation}
\begin{figure}[h]
    \centering
    \includegraphics[width=7.5cm]{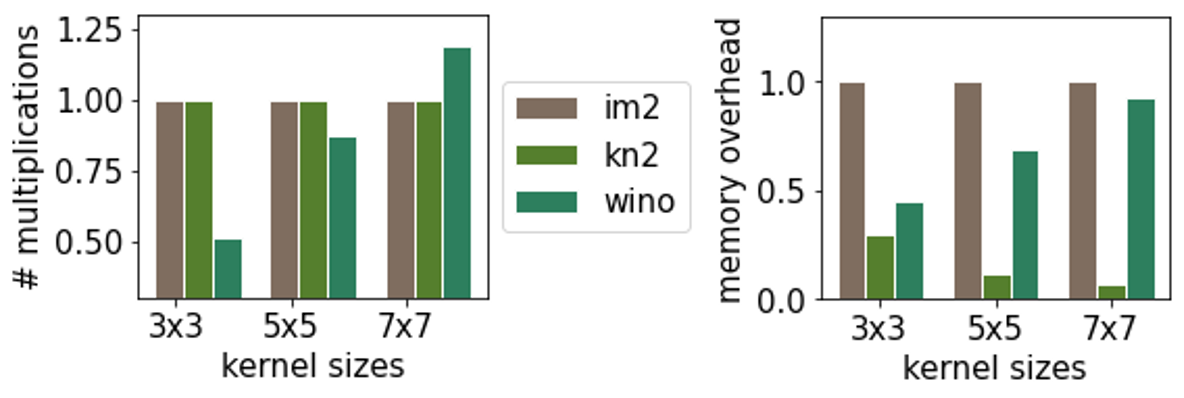}
    \vspace{-2mm}
    \caption{Computation and Memory Loads of GEMM-CONV algorithms on different layer configurations}
    \label{fig:motivation}
\end{figure}

State-of-the-art CNNs \cite{szegedy2015going,szegedy2017inception,ioffe2015batch,szegedy2016rethinking} have highly complex architectures with multiple branches and wide variations in CONV layer configurations. As illustrated in Figure~\ref{fig:motivation}, using three common layer configurations, relative performance of the three GEMM based convolution algorithms depends heavily on the layer configuration, Thus, using a single algorithm across all the layers is not an optimal strategy to minimize the latency of CNN inference. Rather, the ability to switch algorithms between layers is needed. 
Enabling efficient algorithm switching requires us to solve the following problem: (a) As wide variety of CONV configurations and convolution algorithms exist, a unified architecture which is general enough to execute all combinations, while, simultaneously supporting algorithm specific optimizations to extract maximum performance is needed (Section~\ref{sec3}). (b) As each algorithm requires the input and produces the output in its own specific format a low overhead data re-ordering mechanism is needed (Section~\ref{laytrans}). (c) As state-of-the-art-CNNs are extremely deep with number of CONV layers easily surpassing 50 or 100 (Inception-v4 has 141 CONV layers), a low complexity algorithm, which can handle the exponential combinatorial explosion in design choices ($3^{141}$ for Inception-v4) is needed to determine optimal algorithm for each layer (Section~\ref{sec:pbqp}).  

We solve the problems mentioned above and present DYNAMAP --- a framework that takes any CNN model and maps it onto FPGA in order to obtain extremely low latency CNN inference (Section~\ref{sec:dse}).  
\subsection{Related Work}
\cite{podili2017fast,lu2017evaluating} developed efficient FPGA accelerators for Winograd CONV expressed as hadamard multiplication between input tiles. \cite{zeng2018framework,zeng2017fast,niu2020reuse} focus on frequency-domain CNN acceleration which are usually advantageous for large kernels. \cite{zhang2020evaluating} used kn2row and achieved state-of-the-art speedup on Inception v-4 which has more memory-bounded layers. While these works adopt certain strategies to accelerate CNN inference, they do not explore the flexible switching of different algorithms within the same network.

\cite{ye2020hybriddnn} proposed a hls-implemented accelerator that allows switching between Winograd and spatial CONV, achieving high-throughput inference on VGG-16. However, their design is not optimized for more memory-bound CONV layers such as those in Googlenet, mobilenet, Resnet and Inception networks. \cite{incepv2} proposed to combine fast convolution algorithms including Winograd and Fast Fourier Transform by developing separate modules for the two strategies and partition the resource for different layers executing each; \cite{zhang2020dnnexplorer} explores a hardware DSE scheme that allocates resources to a generic module that accelerate majority of the layers and specific modules heavily customized and dedicated to certain types of layers.  While these works exploit fine-grained layer specific tuning, their designs do not maximally re-use the available resource when executing no-batch inference. This is because the dependencies in CNN graph leads to idling of these specific modules when single image inference needs to be performed.
\cite{cong2014combining} proposes a polynomial-time heuristic to a general module selection problem for streaming applications. While this approach focuses on throughput-area tradeoff fitting multiple operators where module circuit implementations are selected(mapped) to each operator, it does not capture CNN-specific algorithm choices such as im2col, kn2row and Winograd to CONV operators(layers). In contrast, we exploit maximal circuit-reuse across different layers and map different algorithms to layers for low-latency CNN inference. To exploit both layer-wise fine-tuning and hardware utilization, we take a different approach than the existing works - We define a unified computing unit that is general enough to execute all layer-algorithm combinations, while, simultaneously supports algorithm switching and algorithm-specific optimizations for maximum performance.

%% file: 3_Architectural_Design.tex
\section{Architecture Overlay design for Dynamic Algorithm Switching}
\label{sec3}
\begin{figure}[h]
    \centering
    \includegraphics[width=8.5cm]{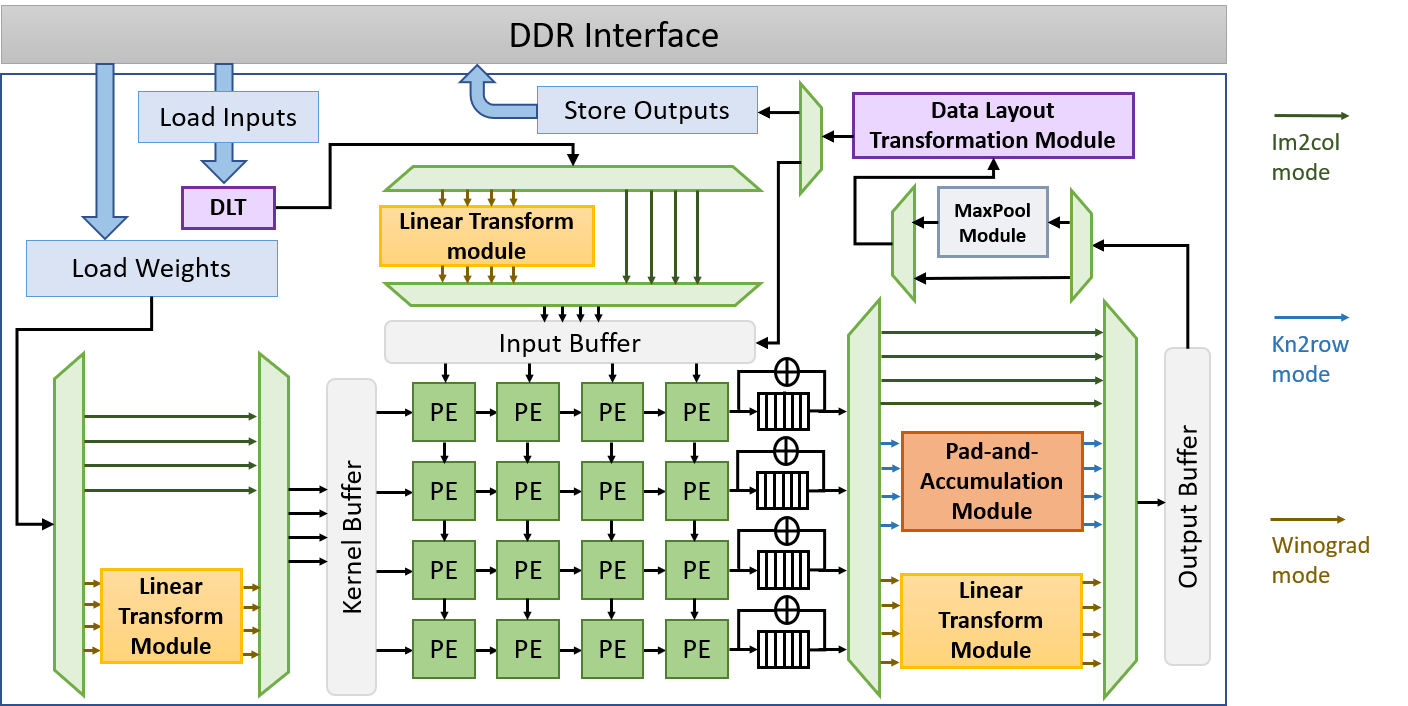}
    \vspace{-2mm}
    \caption{Architecture Overview}
    \label{fig:overview}
\end{figure}
We develop a unified architecture template with a shared central Computing Unit which can be used by all the supported algorithms and separate algorithm-specific auxiliary functional modules. To address the challenges of: (a) hardware under-utilization due to different layer-specific parallelism requirements and (b) algorithm switching overhead, we propose the following optimizations, respectively: (a) a novel PE design that enables stall free computations between layers and enables selection of algorithm specific optimal systolic array dataflow to extract maximum parallelism in GEMM operations (Section~\ref{sec:dfpe}), (b) a low overhead data layout transformation module to enable fast algorithm switching between layers (Section~\ref{laytrans}).
\subsection{Architecture Overview}
Figure \ref{fig:overview} shows a high-level overview of the template architecture design. The accelerator is composed of a GEMM Computing Unit performing, Linear Transform Modules for Winograd, a Pad-and-Accumulation Module for kn2row, Data Layout Transformation (DLT) Modules for data re-ordering when switching between different layers and algorithms, and a Pooling Module for max pooling on spatial feature maps. The Computing Unit (CU) is a $P_{SA1}\times P_{SA2}$ 2-D systolic array of Multiply-Accumulation (MAC) units optimized for GEMM. The Input Buffer and Kernel Buffer are organized into $P_{SA1}$ and $P_{SA2}$ banks, respectively; Each bank stores a partition of input feature / kernel matrix. During GEMM execution, $P_{SA1}$ and $P_{SA2}$ data elements are read concurrently by parallel PEs in the systolic array, and written to $P_{SA1}$ or $P_{SA2}$ banks of output buffer in parallel.

Under im2col mode, the Toeplitz matrices of input feature maps and kernel parameters are loaded into the Input and Kernel Buffers. The output feature maps are directly written into the output buffer. 

Under kn2row mode, the independent $1\times 1$ unit convolutions are expressed as a series of GEMM calls. Then Pad-and-Accumulate Module shifts each intermediate output patch to align with the position determined by the original $K_1 \times K_2$ kernel and produces the final output feature maps using an accumulation buffer. The bank indices and address offsets for Pad-and-Accumulation are pre-computed based on CONV layer meta data \cite{zhang2020evaluating}. The "unit-CONV GEMM" and "Pad-and-Accumulate" phases are pipelined enabling CU to start working on the next patch of unit-CONV GEMM while accumulation buffer still processes the last patch. This reduces the overall "Pad-and-Accumulate" overhead of kn2row.  

Under Winograd mode, both input feature maps and kernels are fed into the Linear Transformation Modules so that the GEMM operates in Winograd-transformed space. Each of $(m+r-1)\times (m+r-1)$ input feature map tile (overlapped by $r-1$ in adjacent tiles) and $r\times r$ kernel tile are transformed into a $(m+r-1)\times (m+r-1)$ sized tile. All the tiles are scattered and reordered into $(m+r-1)\times (m+r-1)$ independent input and kernel matrices \cite{lavin2016fast} sized at $(\frac{H_1}{m}\times \frac{H_1}{m}, C_{in})$ and $C_{in},C_{out}$, respectively. These GEMMs are fed into the systolic array sequentially and the output is again multiplied with transformation matrices to recover the output tensor shape $({O_1}\times {O_2}, C_{out})$. Linear Transform Module requires multiplication by constants and additions which are determined by Winograd hyper-parameters $(m,r)$. For instance, in a $F(2,3)$, the transformation matrices are only composed of values of $\pm 1$ and $\pm \frac{1}{2}$, which can be easily implemented using shift and add operations.

\subsection{Dataflow-switchable \& Stall-free PE design}
\label{sec:dfpe}
Performance of GEMM on a fixed systolic array heavily depends upon the dimension chosen for parallelism. For example, consider a $31 \times 31$ systolic array size for multiplying input matrices of sizes $(a,b), (b,c)=(62,124), (124,64)$. Parallelizing along dimensions $a, c$ and breaking each input matrix into tiles sized at $(124,31)$ and $(124,31)$, respectively, will require extensive zero padding in the last tile which will have only 2 columns along dimension $c$. The effective PE utilization will only be $68\%$. However, if we parallelize along dimension $a,b$ instead, no under-utilization will occur. 

\begin{scriptsize}
\begin{table*}[]
\caption{Tensor Layout transformations}
\begin{tabular}{cccccccccc}
\toprule
& \textbf{I}& \textbf{Step\_b} & \textbf{Step\_d} & \textbf{I1} & \textbf{inc\_b2}& \textbf{inc\_d2}& \textbf{I2} & \textbf{inc\_b3}     & \textbf{inc\_d3}     \\
\midrule
\midrule
\textbf{3D Tensor $\rightarrow$ Toeplitz} &  $O^2$            & $S$              & $K_1K_2$         & $K_1$       & $1$                  & $1$                  & $K_2$       & $H_1S$               & $1$                  \\\midrule
\textbf{3D Tensor $\rightarrow$ Winograd} & $\frac{H_1H_2}{m^2}$ & $m$              & $-\frac{H_1H_2(m+r-1)^2}{m^2}+1$              & $m+r-1$     & $1$                  & $\frac{H_1H_2}{m^2}$ & $m+r-1$     & $H_1$                & $\frac{H_1H_2}{m^2}$ \\\midrule
\textbf{Winograd $\rightarrow$ 3D Tensor} & $\frac{H_1H_2}{m^2}$ & $1$              & $m^2$            & $m$         & $\frac{H_1H_2}{m^2}$ & $1$                  & $m$         & $\frac{H_1H_2}{m^2}$ & $1$                  \\\bottomrule
\end{tabular}
\label{tab:layout}
\end{table*}
\end{scriptsize}

\begin{figure}[ht]
    \centering
    \includegraphics[width=6.5cm]{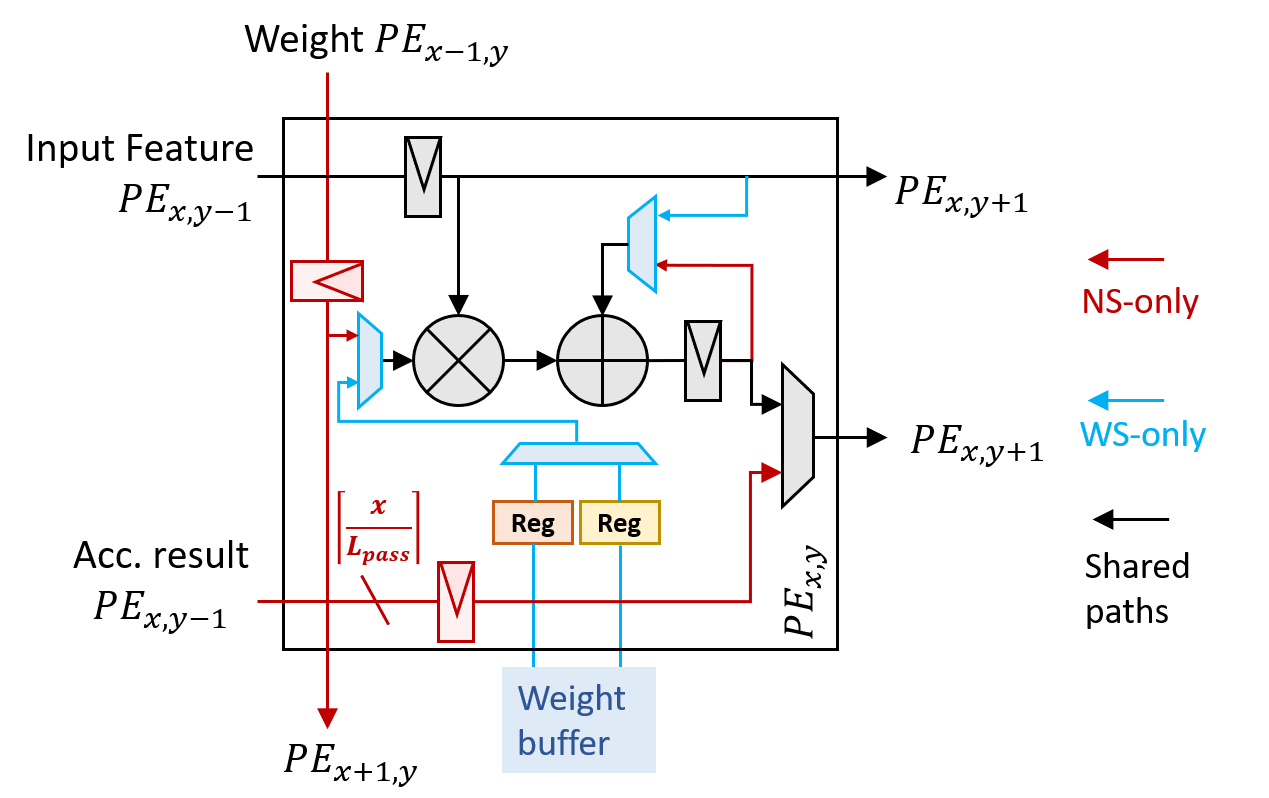}
    \vspace{-2mm}
    \caption{PE dataflow and optimizations to support stall-free operations}
    \label{fig:pe}
\end{figure}

To handle such scenarios, we develop a novel PE design that allows for no-overhead switching between different dataflow to improve PE utilization.
\textbf{Non-stationary (NS) Dataflow:} % (Figure \ref{fig:ns}). 
The input matrices $W(b\times c)$ and $X(a\times b)$ are partitioned along dimensions $a$ (or $c$) into tiles of size $b\times P_{SA1}$ (or $P_{SA2}$). Each Processing Element (PE) computes a vector-product that contributes to one output feature.
In each clock cycle each PE performs one multiply-accumulation (MAC) and shifts the input and weight along two directions. Once all the MAC computations for a pixel are finished, the final result is shifted out of the PE and the PE proceeds to work on a new pixel. The NS datapath in a PE is shown in Figure \ref{fig:pe} with black and red-colored wires. Each pass of matrix partitions incur an initialization overhead $I_{SA}$ that is proportional to $max(P_{SA1},P_{SA2})$ and in a naive implementation will be incurred for each pass in each layer. We alleviate these overheads to implement stall-free GEMM as follows (Figure \ref{fig:pe}): MUX highlighted in grey selects between shifting accumulation result of $PE_{x,y}$ and other PEs. When one PE completes the dot product for one pixel, the accumulation result is directly shifted out, and during the computation of the next PE, accumulation results of other PEs can be shifted concurrently such that $I_{SA}$ is overlapped with next-pass computation.
To further avoid additional stalls due to accumulation result congestion between passes when $b<P_{SA}$ (commonly occurs in layers with small filter and shallow feature maps), we widen the wire(s) used to shift down accumulation results by $\frac{j}{L_{pass}}$ for PEs at the $j^{th}$ row. This ensures that the rate at which outputs are shifted out of each PE matches the rate at which outputs are generated. 
\textbf{Weight-Stationary (WS) Dataflow:} In \textbf{WS}, in each pass, the PEs pre-load a (stationary) block of the weight/kernel matrix sized at $P_{SA1}\times P_{SA2}$ into their local registers. Then the input matrix is fed as tiles of size $P_{SA1}\times a$ into the systolic array in a pipelined fashion. In each clock cycle each PE performs a MAC operation, shifting the input to the next neighbor along $P_{SA1}$ direction and shifting the partial result down to accumulate the partial sums. Each $\frac{b}{P_{SA1}}$ pass produces an intermediate accumulation result of a corresponding $P_{SA2}\times a$-sized partition of the output matrix. This is accumulated at the bottom of the systolic array using accumulators connected with FIFOs of depth $P_{SA1}+c$. After one round (a total of $\frac{b}{P_{SA1}}$ passes) is complete, the final results of one $P_{SA2}\times a$ partition of the output matrix is written back. Such rounds are repeated for every block-tile pairs until the entire GEMM is covered. To remove the initialization overheads in each pass, we modify the basic PE design to pre-load weights (input) in a ping-pong manner using two shift registers (Figure \ref{fig:pe}: highlighted in blue). The $P_{SA1}\times P_{SA2}$ weight block for the next pass is pre-fetched into the register while the current pass is still being processed. \textbf{Input-Stationary (IS) Dataflow} is the mirror of WS: in each pass the PEs pre-load a (stationary) block of the input matrix, shifting the weight and result tiles.

\begin{figure}[h]
    \centering
    \includegraphics[width=8cm]{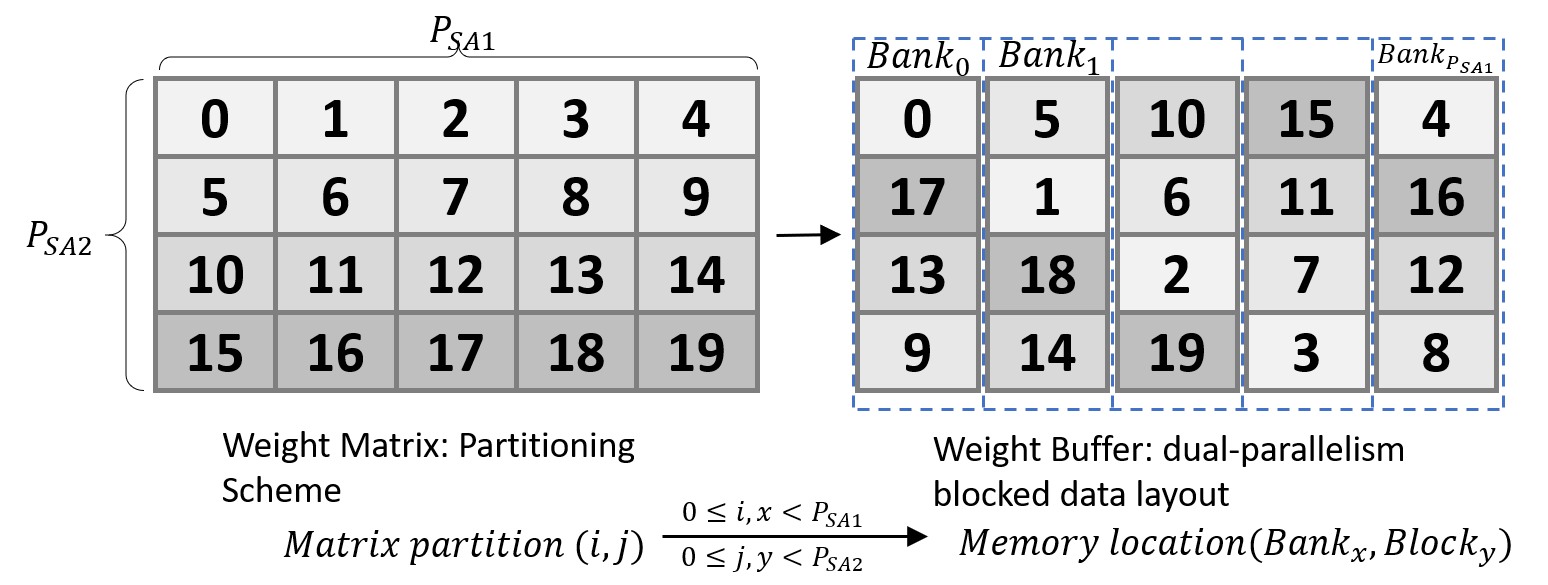}
    \vspace{-2mm}
    \caption{Dual-parallelism blocked data layout}
    \label{fig:dl}
\end{figure}

\textbf{Dual-parallelism Blocked Data Layout:} WS (IS) dataflow accesses input (weight) matrix in a transposed fashion compared to that in NS: In NS, the input (weight) matrix is partitioned along the $b$ dimension, while in WS (IS) it is partitioned along dimension $a$ ($c$). Therefore, even when operating on our stall-free PEs, there will still be bank conflicts which will stall the execution when switching between different dataflows. In practice, if we store weight data into $P_{SA1}\times P_{SA2}$ individual banks, the design is not scalable to lower-precision inferences as the PEs in the systolic array easily outnumber available FPGA on-chip memory block resources. To resolve this problem we use a dual-parallelism blocked data layout as shown in Figure \ref{fig:dl}.
We partition the matrix along both systolic array dimensions and store the blocks in a circular-shifted manner, Equation \ref{dlmap} shows the mapping between feature block $(i,j)$ and its location $(Bank_x,Block_y)$ in on-chip SRAM. Using such bank layout we ensure single-cycle parallel access in both directions can be achieved without bank conflicts.
\begin{footnotesize}
\begin{equation}
\label{dlmap}
\begin{aligned}
&x=(i+j) \% P_{SA1}\\
&y=\left\{\begin{array}{ll}
i \% P_{S A 2} & \text { if }(i+j)<P_{SA1} \\
i-\left(P_{S A I}-P_{S A 2}\right) & \text { if }(i+j) \geqslant P_{SA1}
\end{array}\right.
\end{aligned}
\end{equation}
\end{footnotesize}
\vspace{-2mm}
\subsection{Data Layout Transformation Module}
\label{laytrans}
Each algorithm --- im2col, kn2row and Winograd requires input and produces output in a specific layout. 
We design the Data Layout Transformation (DLT) Module to achieve layout transformation on-the-fly with minimal overheads. At data-store (data-load) side,  DLT module streams in the [on-chip SRAM (DRAM) address, data] tuples, converts the output layout of the previous layer into the correct input tensor layout for the algorithm implemented in the following layer, generates the [DRAM (on-chip SRAM) address, data] tuples and stream out to the DDR controller. As the DLT at data-load side performs symmetric operations as that at data-store side with flipped on-chip SRAM / DRAM address tuples, we only show the transformation scheme for data-store side. 

While all three algorithms have different data layout for the input tensor shape, im2col and kn2row outputs the intermediate feature map in the same layout - spatial 3D tensor layout. Therefore the DLT Module selects from one of six available combinations for layout conversion. When both layers use kn2row algorithm, the output layout is the same (3D tensor) as the next input layout. In this case, the transformation is simply a one-to-one matching between consecutive on-chip SRAM and DRAM addresses. We list the other conversions below:
\begin{figure}[H]
    \centering
    \includegraphics[width=6.5cm]{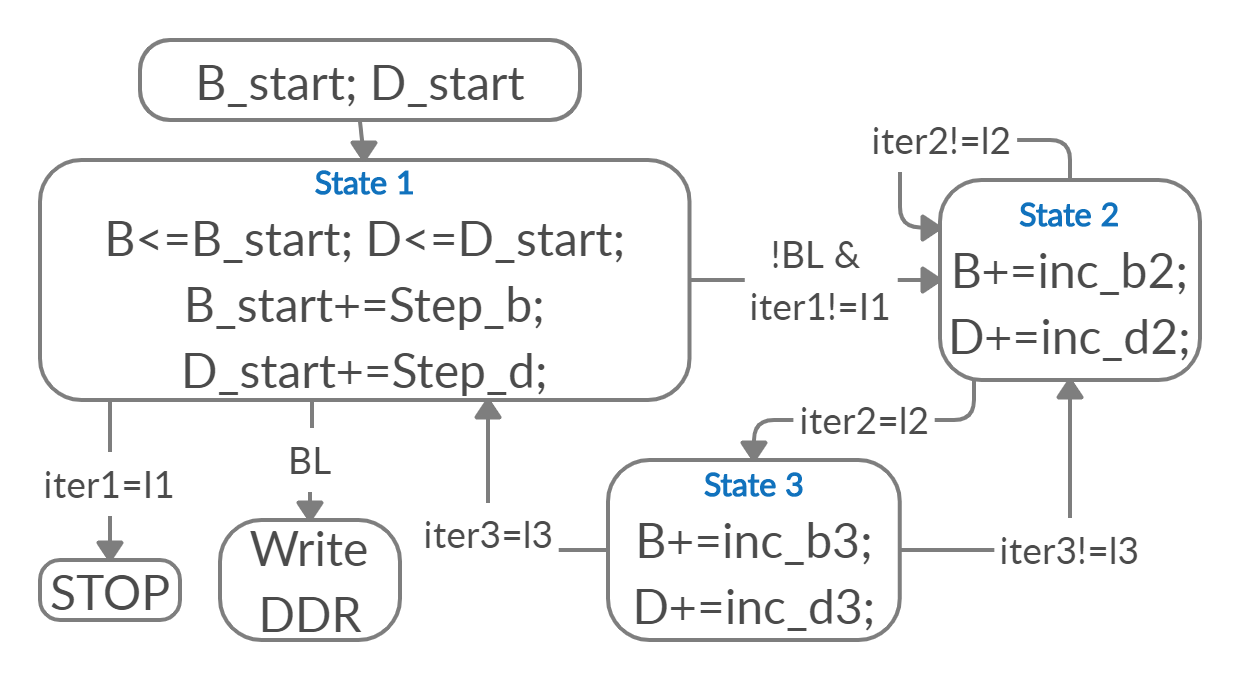}
    \caption{LTU (Data-Store side): FSM flow}
    \label{fig:3d_toep}
\end{figure}
\subsubsection{3D Tensor Layout Transformations} 
The input layout for im2col is known as the Toeplitz matrix, where each group of $C_{in}$ windows of input feature maps corresponding to a filter size is stretched into a column of the Toeplitz matrix, sized at $(O_1O_2,K^2C_{in})$. We define the output layout of im2col and kn2row as spatial 3D Tensor Layout, which is a matrix of shape $(H_1H_2,C_{in})$. Figure \ref{fig:3d_toep} shows a high-level Finite-State Machine diagram of the mechanism adopted in a Layout Transformation Unit (LTU) implementing the transformations, where $B, D$ represent the on-chip SRAM and DDR addresses corresponding to a data point in the feature map. Incrementing $B$ by 1 (or $H*S$) means jumping to the address of the adjacent data (or data at distance of $S$ rows) in the original feature map in on-chip SRAM to obtain a new tuple, and incrementing $D$ by $1$ means setting the next consecutive address as write-back address. The generated (DDR address, data) tuples are buffered until DDR transfer burst length (BL) is saturated, and are then sent to DDR interface.  Iter\_x denotes the counters that keep track of the number of times state x is visited. The process for transforming a 3D Tensor output to a Toeplitz input is shown in Table \ref{tab:layout} first row, state 2 loops inside each row (length $K_1$) of sliding window, state 3 iterates all $K_2$ rows in a sliding window and state 1 steps over all overlapped sliding windows.
\subsubsection{Winograd Layout Transformation} 
In Winograd Input Tensor Layout, $\frac{H_1H_2}{m}^2$ input feature tiles (each sized at $(m+r-1)^2$) are stretched into rows and adjacent tiles share an overlap with width of $r-1$. As each tile sized at $(m+r-1)^2$ is initially scattered to $(m+r-1)^2$ different matrices for separate matrix multiplications \cite{lavin2016fast}, $\frac{H_1H_2}{m}^2$ elements corresponding to the same relative position in each overlapping tile should be adjacent in on-chip SRAM, When transforming from 3D tensor layout to Winograd input Layout (Table \ref{tab:layout} row 2). Consistently, in the Winograd output Tensor Layout, the $m^2$ elements of each output tile are scattered to locations spaced out at $\frac{H_1}{m}$ horizontally or $\frac{H_2}{m}$ vertically, and $\frac{H_1H_2}{m}^2$ elements from different tiles are adjacent in the Output Buffer. Therefore, to transform from Winograd output layout to Toeplitz layout, we first need to restore the 3D Tensor layout (Table \ref{tab:layout} row 3), then transform to Toeplitz input layout using row 2 configurations. To avoid extra roundtrip to DRAM, we double-buffer the Output Buffer into two bank groups, where the systolic array writes to bank group A, LTU \#1 transforms 3D Tensor and writes into bank group B, while LTU \#2 takes input from bank group B and writes into DRAM.

\subsection{Pooling}
MaxPool and AvgPool are the most common Pooling layers. To avoid expensive data traffic between the host processor and FPGA, we allocate dedicated hardware module for MaxPool, and express AvgPool of window $K_1\times K_2$ as equivalent to a 2D convolution with a $K_1\times K_2$ kernel where each element is of value $\frac{1}{K_1\times K_2}$. The hardware building block for the MaxPool module are the Pooling Units (PU).
Each PU contains a Horizontal PU (HPU) to read data from input feature map horizontally in a pipelined manner. The HPU outputs one intermediate Pooling result each clock cycle. After HPU produces $K_1$ rows of intermediate Pooling results, the Vertical PU (VPU) with the same architecture starts processing in the vertical direction, producing one Pooling result each clock cycle, overlapping the HPU in a pipelined fashion. We use an array of PUs to exploit parallelism across feature maps.

%% file: 4_pbqp.tex
\section{Optimization Formulation for Algorithm Mapping}
\label{sec:pbqp}
Given a graph representation of a CNN model and the target FPGA platform, we need to determine the following: (a) For each layer, the choice of the convolution algorithm and the choice of the dataflow. We call this algorithm-dataflow pair.  (b) parameters to customize architecture overlay for the CNN model on the target platform. We discuss (a) in this section, and (b) in Section~\ref{sec:dse}. Note that as optimal dataflow for each algorithm in each layer can be determined in step (b), algorithm mapping here implicitly implies algorithm-dataflow pair mapping.

As described in Section~\ref{laytrans}, the choice of algorithm-dataflow pair not only impacts the execution time of the layer, it also impacts the execution time of all the neighboring layers as represented by the edges in the graph $G$. This is because data layout transformations are needed to ensure data is available to the algorithms in the correct format. Thus, greedily selecting algorithm-dataflow pairs that minimize the execution time at each layer will not minimize the overall execution time of the CNN.

We assume the CNN graph is $G = (V, E, C_v, T_e)$, with each vertex $v \in V$ representing a layer of the model and each edge $e \in E$ representing the ordering between two layers. Let vertices be labelled $\{1,\dots,N\}$, where $N=|V|$. $C_v$ is the cost vector array that represents the computation costs of the vertices (Section~\ref{ssec:costmatrix}) under different algorithm-dataflow pairs. For vertex $i$, $\vec{c}_i$ denotes the cost vector. $T_e$ is the set of transition cost matrices that represent the cost of data layout transformation between vertices (Section~\ref{ssec:txmatrix}). $T_{ij}$ denotes transition matrix for each edge $(i,j)$. The objective is to determine algorithm-dataflow mapping for each layer of CNN such that the cost --- total latency of executing the CNN is minimized. $\vec{x}_i$ is a 0-1 assignment vector with $\vec{x}_i(k) = 1$, if algorithm $k$ is chosen and 0 otherwise. Exactly one entry of $\vec{x}_{i}$ can be set to 1.  The problem can be formulated as follows: 
\begin{equation}\begin{array}{c}
minimize \sum_{1 \leq i<j \leq N} \vec{x}_{i}^{T} T_{i j} \vec{x}_{j}+\sum_{1 \leq i \leq N} \vec{x}_{i}^{T} \vec{c}_{i} \\
\text {s.t.} \\
\vec{x}_{i} \in\{0,1\}^{\left|\vec{c}_{i}\right|} \quad \forall 1 \leq i \leq N \\
||\vec{x}_{i}||_{1} == 1 
% \quad \quad \vec{x}_{i}^{T} \overrightarrow{1}=1
\end{array}\end{equation}

This problem formulation is known as Partitioned Boolean Quadratic Programming (PBQP) problem~\cite{scholz2002register}. PBQP has been used to model a number of problems in compiler optimization such as register allocation for architectures with irregular instruction sets \cite{scholz2002register}, and instruction selection on DAGs~\cite{eckstein2003code}. 

PBQP is NP-Complete~\cite{scholz2002register,anderson2018optimal}. However, we show that for a class of graphs, known as \textit{series-parallel graphs}, PBQP can be solved in polynomial time. Moreover, we show that the graphs of a majority of popular CNN architectures fall into this class. This allows us to develop a polynomial time optimal algorithm for the algorithm mapping optimization problem defined above.  

\textbf{Definition 1:} A (undirected) graph, with two distinguished vertices -- source $s$ and sink $t$, is a series-parallel graph if it can be turned into a $K_2$ graph (a graph with two vertices connected with an edge) by a sequence of the following operations~\cite{duffin1965topology}:
\begin{enumerate}
    \item Remove a degree 2 vertex other than $s$ or $t$ and the edges incident on it. Directly connect the two neighbors with a single edge.
    \item Replace a pair of parallel edges with a single edge that connects the two endpoint vertices.
\end{enumerate}
\begin{theorem}
\label{thm1}
PBQP can be solved in polynomial time if the graph is a series parallel graph. Moreover, for a graph with $N$ vertices and $d = \max_i \left|\vec{c}_{i}\right|$, the running time is $O(Nd^2)$.
\end{theorem}
\begin{theorem}
\label{thm}
On any series-parallel graph, reduction operations (1) and (2) preserves the optimality of PBQP.
\end{theorem}
\vspace{-2mm}
\begin{proof}
We prove Theorem \ref{thm1} and \ref{thm} by induction. 

The \textbf{inductive property} is that any series-parallel graph can be constructed from a base K-2 Graph $(s,t)$, where $s$ is its source and $t$ is its sink, by a combination of the following two steps:

\textbf{Base steps:} (1) adding a vertex with an edge connected to $s$ or $t$ of the base K-2 Graph; (2) adding an edge between $s$ and $t$ of the base K-2 Graph;

\textbf{Inductive steps:} (1) adding a vertex with an edge connected to the source or sink of an existing series-parallel graph; (2) adding an edge between the source and sink of an existing series-parallel graph; Note that adding a degree-2 vertex connected to the source and sink of a K-2(or, series-parallel) graph $G_s$ is equivalent to performing Base(or, Inductive) step (1) followed by Inductive step (2) on $G_s$.

In the following, we prove that any series-parallel graoh constructed this way can be reduced to a K-2 graph with preserved optimality.

\textbf{Base Cases: (1)} For a graph $G^1$ with three vertices $i,j,k$ and edges $(i,k), (k,j)$, the optimal solution is obtained by choosing algorithms $(d_k)$ for vertex $k$ such that $d_k=argmin\{\vec{c}_i(d_i) + T_{i,k}(d_i, d_k) + \vec{c}_k(d_k) + T_{k,j}(d_k, d_j) + \vec{c}_j(d_j)\}$ for all $(d_i, d_j)$ pairs in $G^1$.
Applying operation 1 on vertex $k$ to obtain $K_2$ graph --- $G^1_{K_2}$ and setting $T_{ij}(d_i, d_j) =$ $ \min_{d_k}\{T_{i,k}(d_i, d_k) + \vec{c}_k(d_k) + T_{k,j}(d_k, d_j) \}$ for all $(d_i, d_j)$  in $G^1_{K_2}$, the optimal solution on the reduced graph can be found by iterating through $(\vec{c}_i(d_i), \vec{c}_i(d_j))$ pairs. Thus, optimality is preserved. 
\textbf{(2)} For a graph $G^1$ with two vertices $i,j$ and two parallel edges between vertices $(i,j)^1$ and $(i,j)^2$, the optimal solution is $min\{\vec{c}_i(d_i) + T^1_{i,j}(d_i, d_j)  + T^2_{i,j}(d_i, d_j) + \vec{c}_j(d_j)\}$, where $T^1$ and $T^2$ are matrices of the parallel edges.
Applying operation 2 by updating $T_{i,j}(d_i, d_j) = T^1_{i,j}(d_i, d_j) + T^2_{i,j}(d_i, d_j)$, the optimality is preserved.

\textbf{Inductive Hypothesis: } We assume that a subgraph, $G^S=(V^S,E^S)$, of a series-parallel graph, where $|V^S|=N,|E^S|=M$, can reduce to a $K_2$ graph with preserved optimality.

\textbf{Inductive Cases:} We show that a series-parallel graph $G=(V,E)$ constructed by adding a vertex/edge to $G^S$ specified above can still reduce to a $K_2$ graph with optimality preserved: \textbf{(1)} If $G(|V|=N+1,|E|=M+1)$ is constructed by adding a vertex, $k$, with an edge connected to the source $s$ or the sink $t$ of $G^S$ by an edge $(k,s)$ or $(t,k)$, the optimal solution can be obtained by treating $G^S$ as its reduced $K_2$ graph (due to Inductive Hypothesis). Assuming we connected $k$ to $t$, $G$ can be treated as having three vertices $s,t,k$ connected by two edges $(s,t),(t,k)$, which is reducable to $K_2$ graph with $s,k$ connected by an edge using the same method as shown in Base case: (1). \textbf{(2)} If $G(|V|=N,|E|=M+1)$ is constructed by adding an edge, $(s,t)$, connecting the source $s$ and the sink $t$ of $G^S$, the optimal solution can be obtained by treating $G^S$ as its reduced $K_2$ graph (due to Inductive Hypothesis), therefore $G$ can be treated as having two vertices $(s,t)$ and two parallel edges, which is reducable to $K_2$ graph using the method shown in Base case: (2). Therefore any series-parallel graph can be reduced to a $K_2$ graph, preserving the optimality of the solution on the original graph. As described in the base cases, each reduction operation (1) or (2) requires $O(d^2)$ amount of work and is performed $O(N)$ times in total for a given graph.
\end{proof}

\begin{figure}[ht]
    \centering
    \includegraphics[width=8.5cm]{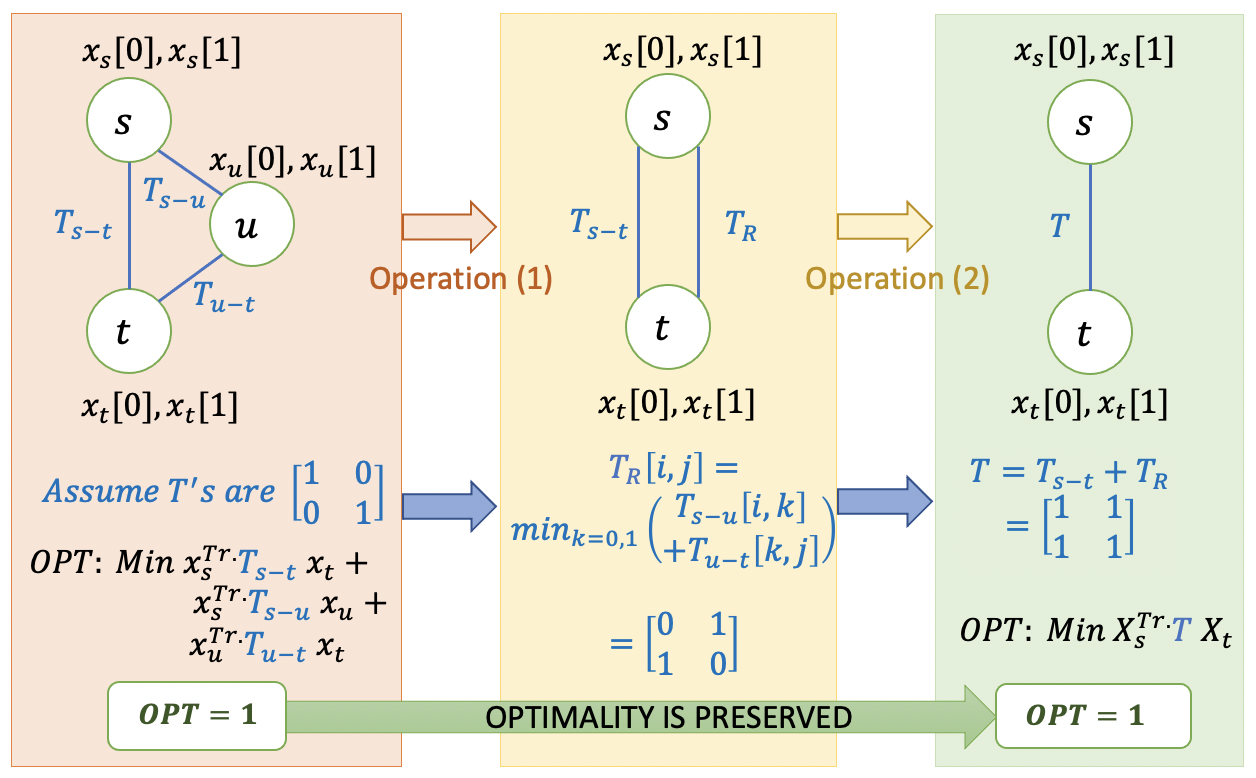}
    \vspace{-2mm}
    \caption{Sample Reduction Process}
    \label{fig:samplered}
\end{figure}

We demonstrate the proof for the optimality-preserving reduction with a simple example in Figure \ref{fig:samplered}, where we assume $N=3, d=2$ and $c_i=0, 0\leq i\leq d$.

\begin{lemma}
Resnet~\cite{resnet}, VGG~\cite{vgg}, Alexnet~\cite{alexnet}, 
etc. which do not have any branches are series parallel graphs.
\end{lemma}
\vspace{-2mm}
\begin{proof}
Let the input layer be denoted by vertex $s$ and the output layer by vertex $t$. The degrees of vertices corresponding to all the other layers for VGG and Alexnet are 2. By repeatedly applying operation 1, we obtain a $K_2$ graph. In ResNet, some vertices have higher degrees due to skip connections. However, the vertices between the end points of each skip connection have degree 2. Thus, repeatedly applying operation 1 on these nodes until an edge parallel to the skip connection edge is obtained, and then applying operation 2 results in a graph with all vertices, except $s$ and $t$, having degree 2. By repeatedly applying operation 1, we obtain a $K_2$ graph.
\end{proof}

\begin{lemma}
GoogleNet~\cite{szegedy2015going}, Inception-v1 to v4 and Inception-ResNet-v1 \cite{szegedy2017inception,szegedy2016rethinking},  which are composed of inception modules are series parallel graphs.
\end{lemma}
\vspace{-2mm}
\begin{proof}
Due to space limitations, we prove this lemma only for Inception-v4. Similar arguments can be made for the other networks. Consider Inception-C block (Figure 6 in~\cite{szegedy2016inception}). The output of the \textit{Filter concat} layer at the bottom splits into 4 branches. The left 2 branches have only degree 2 nodes and can be converted to a single edge by operation 1. For the third branch from left, applying operation 1 on $1 \times 3$ and $3 \times 1$ CONV layers, followed by operation 2 on the resulting parallel edge and then applying operation 1 on $1 \times 1$ CONV layer will result in a single edge. Similarly, the rightmost branch can also be converted into a single edge. The four parallel edges can be merged (operation 2) resulting in a $K_2$ graph. In a similar manner, Inception-A, B and Stem modules can also be reduced to $K_2$. Thus, Inception-v4 network (Figure 9 in~\cite{szegedy2016inception}) can be reduced into a number of $K_2$ graphs connected in series which can in turn be trivially reduced to $K_2$. 
\end{proof}

%% file: 5_DSE.tex
\section{DYNAMAP: Framework Specification}
\label{sec:dse}
\begin{figure}[ht]
    \centering
    \includegraphics[width=7cm]{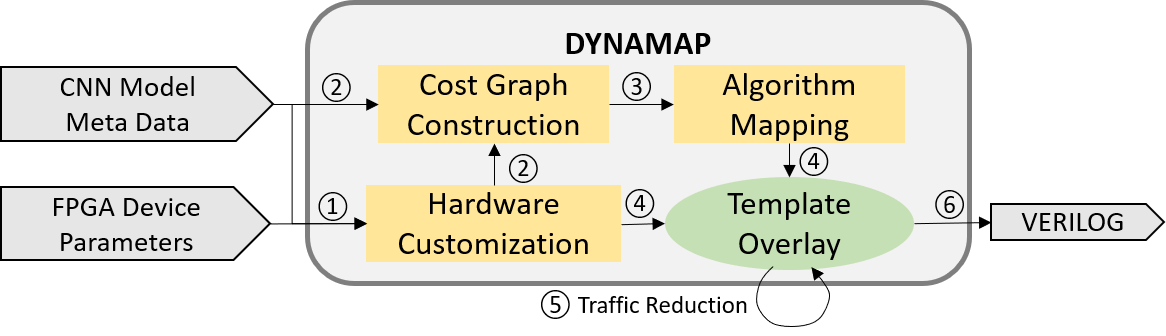}
    \vspace{-2mm}
    \caption{DYNAMAP Software Tool Flow}
    \label{fig:allflow}
\end{figure}
\textbf{DYNAMAP Software Execution Flow:}
DYNAMAP uses a hardware overlay template (Section \ref{sec3}), which is parameterized with $P_{SA1},P_{SA2}$, sequence of $\psi$, and sequence of control signals encoded by specific algorithm mapping that defines the behavior of DLT, Linear Transform and Pad-and-Accumulate modules. As shown in Figure \ref{fig:allflow}, after inputs (FPGA device capabilities, CNN meta data) are provided, DYNAMAP executes the following steps: \circled{1}Algorithm \ref{algo: hardware_algo} first identifies $P_{SA1},P_{SA2}$ and best $\psi$ associated with available algorithms in each layer. \circled{2} These parameters are used to construct and populate the CNN graph as discussed in Section \ref{sec4}. \circled{3} Then, an off-the-shelf PBQP solver \cite{pbqpsolve} is utilized to perform the node reduction steps for algorithm mapping as described in Section \ref{sec:pbqp}. The PBQP solver outputs the optimal algorithm assignment vectors for all layers. \circled{4} Based on the algorithm-dataflow mappings, the template overlay is customized. \circled{5} CNN is scanned to identify any consecutive layers whose total memory consumption do not exceed on-chip SRAM capability. Store-side LTUs are allocated and customized to generate SRAM addresses and store the layer output into the Input Buffer. Thus, on FPGA devices with larger on-chip memory, redundant off-chip data traffic will be avoided. \circled{6} Integrating all the optimizations, control signal sequences are generated to support the algorithm switching on the hardware overlay. The output of DYNAMAP is synthesizable VERILOG program that can be deployed on the target FPGA.

In the following, we discuss in detail (i) the CNN Cost Graph Construction, which assumes a fixed systolic array of size $P_{SA1} \times  P_{SA2}$, and identifies the cost vectors $c$ and Transition matrices $T$ for each layer; (ii) the Hardware Customization, which performs DSE to identify the systolic array dimensions and dataflow mapping to algorithms, providing input to the Cost Graph Construction.

\subsection{Cost Graph Construction}
\label{sec4}

To construct cost graph $G = (V', E', C_v, T_e)$ from 
CNN graph $G=(V, E)$, for each vertex $v^i \in V$, we add a node $v^i_c$ into $V'$. Moreover, for each $v^i \in V \; |\; outdegree(v^i) > 1$, we add another node $v^i_s$ into $V'$
because: Layer $i$ that is connected directly to multiple downstream layers can store the output in only one format. The data load time of each downstream layer is dependent upon this format. The vertex $v^i_s$ is used to capture the format in which layer $i$ needs to store the data to DRAM.   
Now, for an edge $(v^i, v^j) \in E$, if $outdegree(v^i) \leq 1$, we simply add the edge $(v^i_c, v^j_c)$ to $E'$. Else we add the following new edges: $(v^i_c, v^i_s)$ and $(v^i_s,v^j_c) \; \forall j$. 

\subsubsection{Cost Vector Array Construction}
\label{ssec:costmatrix}
Let $\psi$ denote the dataflow. For a GEMM operation with dimensions $a\times b$ (input) and $b\times c$ (weight), the execution time on the systolic array $P_{SA1} \times P_{SA2}$ is given by the following equations: 
\begin{footnotesize}
\begin{equation}
\label{obj_mm}
        \underset{(P_{SA1},P_{SA2},\Psi)}{\operatorname{\mathbb{C}_{mm}}}(a,b,c)= 
        \left\{ \begin{array}{lcl}
        \Psi=NS: \lceil \frac{a}{P_{SA1}} \rceil \times \lceil \frac{c}{P_{SA2}} \rceil \times b + I_{SA}, \\ 
        \Psi=WS: \lceil \frac{b}{P_{SA1}} \rceil \times \lceil \frac{c}{P_{SA2}} \rceil \times a + I_{SA},\\
        \Psi=IS: \lceil \frac{b}{P_{SA1}} \rceil \times \lceil \frac{a}{P_{SA2}} \rceil \times c + I_{SA}
        \end{array}\right.
\end{equation}
\end{footnotesize}
where $I_{SA}$ represents the one-time initialization overhead.

Thus, the latencies of executing a CONV layer on a device with frequency FREQ are given by Equation \ref{im2col_c} for im2col, \ref{kn2row_c} for kn2row and \ref{wino_c} for Winograd $(m,r)$ algorithm.

\begin{footnotesize}
\begin{equation}
\label{im2col_c}
    \underset{(P_{SA1},P_{SA2},\Psi)}{\operatorname{\mathbb{C}_{mm}}}(O_1O_2,K_1K_2C_{in},C_{out})/FREQ
\end{equation}
\begin{equation}
\label{kn2row_c}
    \underset{(P_{SA1},P_{SA2},\Psi)}{\operatorname{\mathbb{C}_{mm}}}(O_1O_2,C_{in},C_{out})\times K_1K_2/FREQ
\end{equation}
\begin{equation}
\label{wino_c}
    \underset{(P_{SA1},P_{SA2},\Psi)}{\operatorname{\mathbb{C}_{mm}}}(\frac{H_1H_2}{m^2},C_{in},C_{out}) +LT )(m+r-1)^2\frac{K_1K_2}{r^2}/FREQ
\end{equation}
\end{footnotesize}
where $LT$ denotes the linear transformation overhead.

\textbf{Cost Vector Array Construction:} Let the number of algorithm-dataflow pairs in $i^{th}$ layer be $|A_i|$. We define the cost vector array $C_v$ consisting of $|V'|$ vectors as follows: For $i \in V_c$, define $\Vec{c}_i \in R^{|A_i|}$, 
% with entry $j$ - $c_{ij}$ 
each entry computed by plugging in the dimensions of layer $i$ into one of the Equations~\ref{im2col_c}-\ref{wino_c} with appropriate $\psi$ in the algorithm-dataflow pair $j$ (Section \ref{sec:custo}).  For $i \in V_s$, define a zero vector $\Vec{c}_i \in R^{\sum_{d'=1}^d|A_{d'}|}$, where $d$ is the outdegree of $i$ and $d'$ is the index for each downstream layer. 

\subsubsection{Transition Matrix Construction}
\label{ssec:txmatrix}
Each layer fetches the input data from the external memory (DRAM), stores it in the on-chip memory, performs computations, and stores the output back into the external memory. Each CONV algorithm has specific input and output layouts. We denote the layout associated with a certain layer $i$ as algorithm format ($AF_i$).
Thus, the transition overhead between two layers is composed of the following: (a) \textbf{Store:} The latency to store the current-layer output from on-chip SRAM into DRAM in the format needed by the next-layer's algorithm;
(b) \textbf{Load:} The latency to load the next-layer input from DRAM into on-chip SRAM in the format needed by next layer's algorithm;
(c) \textbf{Overheads:} The latency to perform calculations such as max pooling, etc. 
Feature map Load/Store latencies can be calculated using the equations shown in Table~\ref{tab:store}. $BW$ denotes the available DDR bandwidth.  For store, $AF_i \rightarrow AF_{i+1}$ means layer $i$ computes using algorithm $AF_i$ and the output needs to be stored (layout transformed) into the format of $AF_{i+1}$.  For load, it means the output of layer $i$ was stored in the format of algorithm $AF_i$ and layer $i+1$ will use the $AF_{i+1}$ as input. 
\begin{scriptsize}
\begin{table}[ht]
\caption{Load/Store Latency}
\vspace{-2mm}
\begin{threeparttable}[t]
\begin{tabular}{cc}
\toprule
\begin{tabular}[c]{@{}l@{}}\textbf{Algo. Format:}\\ $AF_i\rightarrow AF_{i+1}$\end{tabular}                                & \textbf{Load/Store Latency }                                    \\ \midrule\midrule
\begin{tabular}[c]{@{}l@{}}$im2col\rightarrow im2col$\\ $kn2row\rightarrow im2col$\end{tabular}                                & $\frac{O_1O_2K_1K_2C_{out(i)}}{BW}$                     \\ \midrule
\begin{tabular}[c]{@{}l@{}}$im2col\rightarrow kn2row$\\ $kn2row\rightarrow kn2row$\\ $winograd\rightarrow kn2row$\end{tabular} & $\frac{H_1H_2C_{out(i)}}{BW}$                           \\ \midrule
\begin{tabular}[c]{@{}l@{}}$im2col\rightarrow Winograd$\\ $kn2row\rightarrow Winograd$\end{tabular}                            & $\frac{H_1H_2(m+r-1)^2C_{out(i)}}{m^2f(BW,C_{out(i)})}$ \\ \midrule
$Winograd\rightarrow Winograd$                                                                                                 & $\frac{H_1H_2(m+r-1)^2C_{out(i)}}{m^2BW}$               \\ \midrule
$Winograd\rightarrow im2col$                                                                                                   & $\frac{O_1O_2K_1K_2C_{out(i)}}{BW}+ovhd$       \\ \bottomrule
\end{tabular}
\begin{tablenotes}
\begin{scriptsize}
\item[$\ast$]$H_1,H_2,K_1,K_2,O_1,O_2$ are $Layer_{i+1}$ meta data.
\end{scriptsize}
\end{tablenotes}

\end{threeparttable}
\label{tab:store}
\end{table}
\end{scriptsize}

The $1^{st}$ row of Table \ref{tab:store} shows the transformation from 3D Tensor to Toeplitz layout, which incurs some data copies due to overlapping sliding windows but can be streamed out, as consecutive DRAM addresses are accessed (Section \ref{laytrans}). In the $2^{nd}$ row, between $im2col/kn2row$ output and $kn2row$ 3D Tensor input, one-to-one matching is required. With $Winograd$ output features, some re-ordering is required but the amount of data is not changed.
For 3D Tensor to Winograd input layout transformation shown in $3^{rd}$ row, both data re-ordering and data duplication are needed, and the generated DDR addresses are $\frac{H_1H_2}{m^2}$ apart. Note that in Section \ref{laytrans} we show the transformation of feature map with depth $1$, but in practice we access $C_{out(i)}$ altogether for each address increment. Thus, depending on whether each transaction of $C_{out(i)}$ addresses saturate the entire DDR burst length,
burst length wastage may occur.
 We use $f$ to capture such possible wastage of bandwidth:
 \begin{footnotesize}
\begin{equation}f\left(B W, C_{\text {out}(i)}\right)=\left\{\begin{array}{c}
B W \text { if } C_{\text {out}(i)} \geq B L \\
\frac{C_{\text {out}(i)}}{{C_{\text {out}(i)}}+\frac{m^{2}}{H_{1} H_{2}}} \times BW \text { , otherwise }
\end{array}\right.\end{equation}
\end{footnotesize}
The $4^{th}$ row models the transformation from Winograd output to Winograd input layout, taking advantage of the fact that both are in the "scattered" layout, streaming access can be achieved. 
The $5^{th}$ row models the time for the 2-step transformation: Winograd output to 3D Tensor followed by 3D Tensor to Toeplitz layout. We use 2 pipelined LTU operating on double-buffered SRAM, and use $ovhd$ to denote the initialization overhead.

\begin{figure}[ht]
    \centering
    \includegraphics[width=8cm]{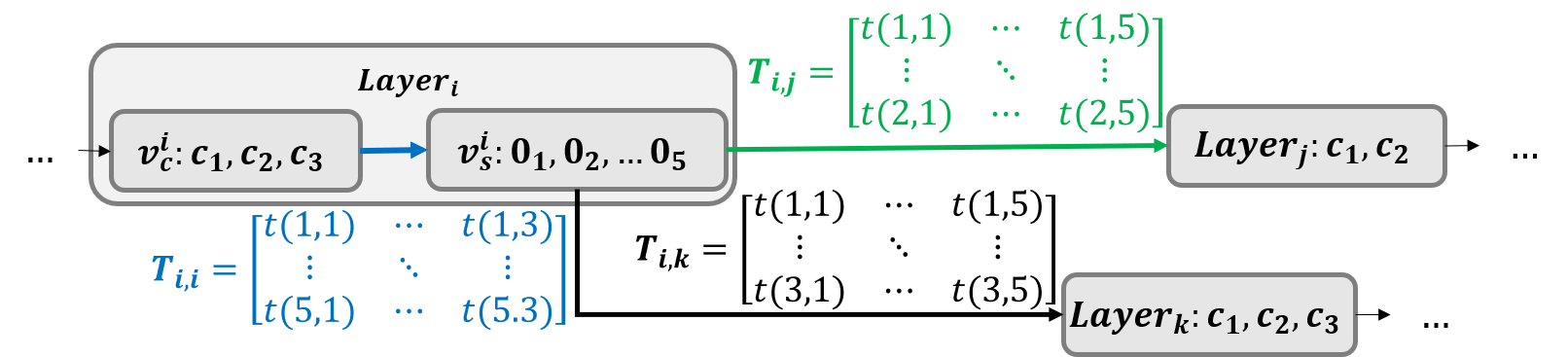}
    \vspace{-2mm}
    \caption{A snippet of an example $\mathcal{G}_{P_{SA1}, P_{SA2}}$}
    \label{fig:dag}
\end{figure}

\textbf{Transition Matrix Construction:} For an edge $(v^i_c,v^j_c)$, the transition matrix $T_{ij}$ of size $|A_i| \times |A_j|$ can be constructed as follows: $T_{ij}(m,n)$ for algorithm $m$ in layer $i$ and algorithm $n$ in layer $j$, entry $T_{ij}(m,n) =Store(m,n,dim(j)) + Load(n,n,dim(j)) $ and other overheads (pooling, etc.) if applicable. 
For an edge $(v^i_c,v^i_s)$ such that $v^i_s$ has $b$ outgoing neighbor vertices, which for simplicity are indexed $\{1,\dots,b\}$, the transition matrix $T_{ii}$ will be of size $|A_i| \times \sum_{b'=1}^b|A_{b'}|$ and can be constructed as follows: $T_{ii}(m;b',o) = Store(m,o,dim(b'))$ and other overheads for algorithm $m$ in layer $i$, storage format corresponding to algorithm $o$ in layer corresponding to vertex $b'$. For an edge $(v^i_s,v^j)$, where $v^j$ is the $b'^{th}$ neighbor of $v^i_s$, 
the transition matrix $T_{ij}$ will be of size $\sum_{b'=1}^b|A_{b'}| \times |A_j|$ and can be constructed as follows: $T_{ij}(o;b',p) = Load(o, p, dim(j))$ where $p$ is the algorithm in layer $j$. 

Figure~\ref{fig:dag} shows a snippet of an example graph $\mathcal{G}_{P_{SA1}, P_{SA2}}$.
\vspace{-2mm}
\subsection{Hardware Customization}
\label{sec:custo}
We use Algorithm~\ref{algo: hardware_algo} to determine: (1) $P_{SA1} \times  P_{SA2}$ and (2) optimal dataflow mapping for each algorithm for each layer. $P_{SA1} \times  P_{SA2}$ is then used to construct the CNN cost graph $\mathcal{G}_{P_{SA1}, P_{SA2}}$. The key idea of Algorithm~\ref{algo: hardware_algo} is as follows: We iterate through possible values of $P_{SA1}$ and $P_{SA2}$ (line 4). For a fixed $P_{SA1}, P_{SA2}$ pair, we calculate the value of empirical total node cost, $\tau_{temp}$, which is the sum of the execution times of all the algorithms over all the layers (line 6-11). Execution time of an algorithm for a layer is calculated using the dataflow that leads to the minimum value (line 7-8). $P_{SA1}, P_{SA2}$ with minimum $\tau_{temp}$ is output.
\begin{footnotesize}
\begin{algorithm}
\caption{Architecture Parameter Identification}
\label{algo: hardware_algo}
\begin{algorithmic}[1]
\renewcommand{\algorithmicrequire}{\textbf{Input:}}
\renewcommand{\algorithmicensure}{\textbf{Output:}}

\State $\tau_{min}\gets \infty${\color{blue}\Comment{Empirical minimum total node costs}}
\State $argminP_{SA1},argminP_{S21}\gets 0$ {\color{blue}\Comment{Systolic Array size mapping}}
\State $\psi[l,a]\gets \emptyset$ {\color{blue}\Comment{OPT. dataflow mapping to layer-algorithm}}
\For {all $(P_{SA1}, P_{SA2})$ s.t. $C(P_{SA1}, P_{SA2}|r)<C_{FPGA|r}$ }
    \State $\tau_{emp}\gets 0${\color{blue}\Comment{Empirical total node costs to be minimized}}
    \For {layer $l=1...L$}
        \For {algorithm $a \in$ [All available algorithms for $l$]}
            \State $c_1,c_2,c_3\gets$ exe. time using NS,WS,IS {\color{blue}\Comment{Equation \ref{obj_mm}}}
            \State $c[l,a]\gets min(c_1,c_2.c_3)$, $\psi[l,a]\gets argmin(c[l,a])$
        \EndFor
        \State $\tau_{emp}\gets \tau_{emp}+sum(c[l,a]) \forall a$
    \EndFor
    \If{ $\tau_{emp} < \tau_{min}$ }
        \State $\tau_{min} \gets \tau_{emp}$
        \State $argminP_{SA1},argminP_{SA2}\gets P_{SA1}, P_{SA2}$ 
    \EndIf
\EndFor
\Ensure OPT. matching: $l,a \gets \psi$, 
\Ensure parameters: $argminP_{SA1},argminP_{SA2}$
\end{algorithmic}
\end{algorithm}
\end{footnotesize}

\begin{figure*}[ht]
    \centering
    \includegraphics[width=17cm]{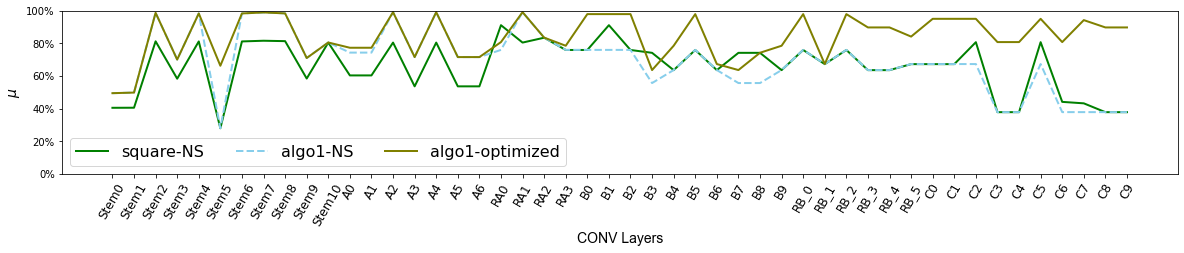}
    \vspace{-2mm}
    \caption{Effective PE utilization under different hardware configurations: Inception-V4}
    \label{fig:opt1_i4}
\end{figure*}
\begin{figure*}[ht]
    \centering
    \includegraphics[width=17cm]{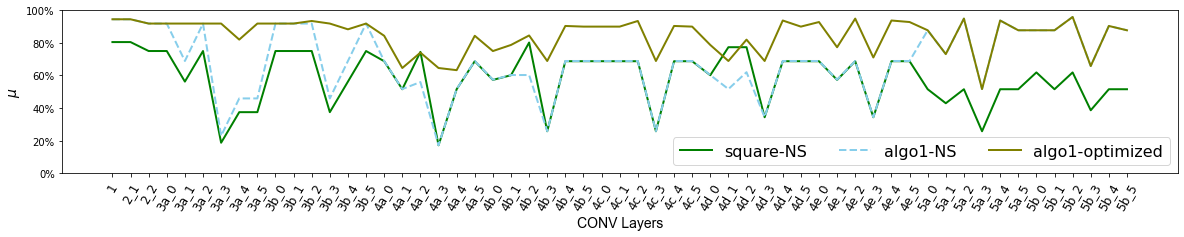}
    \vspace{-2mm}
    \caption{Effective PE utilization under different hardware configurations: GoogleNet}
    \label{fig:opt1_gn}
\end{figure*}

%% file: 6_Experiments.tex
\vspace{-3mm}
\section{Evaluations}
The fundamental objective of our framework is to reduce hardware under-utilization induced by diverse layer shapes and minimize the total end-to-end inference latency. 
In this section, we use our framework to generate the hardware-algorithm co-designs for two state-of-the-art CNNs, GoogleNet\cite{szegedy2015going} and Inception-v4 \cite{szegedy2015going} and show: 
(1) how our dynamic dataflow selection and architecture configuration technique achieves local optimal acceleration at each layer by driving up effective PE utilization; (2) how our novel algorithm mapping achieves global optimal acceleration by improving end-to-end inference latency. 

We use Xilinx Alveo U200 FPGA board hosted on a Xeon Server CPU (E5-2698 v4 @2.2GHz) to evaluate the designs generated by our framework. We use 8-bit fixed-point data representation to perform CNN inference. The designs were synthesized using Vivado 2018. We input the CNN model and FPGA device meta data into our framework to obtain the architecture customization as output. We limit the systolic array DSP consumption to 6084 instead of using all the available DSPs to obtain a fair performance comparison with the state-of-the-art implementations. DYNAMAP returns optimal $(P_{SA1},P_{SA2})$ as (92,66) for GoogleNet, and (95,64) for Inception-V4. The resource utilization are shown in Table \ref{tab: exp comparison}. 

\begin{footnotesize}
\begin{table*}[!ht]
\caption{Comparison of with state-of-the-art implementations}
\vspace{-2mm}
    \centering
    \begin{tabular}{rcccccc}
        \toprule
        & \multicolumn{2}{c}{This Paper} &  \multicolumn{1}{c}{\cite{gn_0}}&\multicolumn{1}{c}{\cite{gn_1}} & \multicolumn{1}{c}{\cite{incepv4}} & \multicolumn{1}{c}{\cite{incep_v4_2}}\\
        
          \cmidrule(lr){2-3}\cmidrule(lr){4-4}\cmidrule(lr){5-5}\cmidrule(lr){6-6}\cmidrule(lr){7-7}& GoogleNet & Inception-v4 & GoogleNet& GoogleNet& Inception-v4 & Inception-v4  \\
        \midrule
        \midrule
        Device & \multicolumn{2}{c}{Alveo U200} & Stratix 10 GX & KU115 & XCVU9P & XCVU9P \\
        Datatype & INT8 & INT8 & INT16 & INT16 & INT8 &INT8\\
        Frequency [MHz] & 286 & 286 & 300 & 250 & 300 & 180  \\
        DSP (\% total) & 6239 (91\%) & 6230 (91\%) & 6304 (55\%) & 4214 (76\%) & 5254 (75\%) & 5130 (75\%)  \\
        On-chip Memory [BRAM/M20K] (\% total) & 2K (93\%) & 2.1K (97\%) & 1949 (17\%) & 2160 (100\%) & 1664 (77\%) & 562 (26\%), 845 (88\%) URAM \\
        LUT/ALM (\% total)& 745K (60\%) & 806K (65\%) & 528K (97\%) & 663K (71\%) & 469K (40\%) & 543K (46\%) \\
        
        \midrule
        Throughput [GOPS/s] & 3568 & 3650 & 557 & 1630 & 3448 &  1528 \\
        Latency/image [ms] & 1.34 & 4.39 & 5.7 & 3.8  & 5.29 & 6.03  \\
        \bottomrule
    \end{tabular}
    \label{tab: exp comparison}
\end{table*}
\end{footnotesize}
\vspace{-2mm}
\subsection{Evaluation of Optimizations}
\subsubsection{Hardware Utilization under different accelerator configurations}
\label{utilexpsec}
 
We define the metric, effective PE utilization $\mu$, as ratio of the total number of effective computations and the total number of computation performed by all PEs. That is,
\begin{footnotesize}
\begin{equation}
    \mu_{i}=\frac{\sum_{t=1}^{T} P E_{\mathrm{on}_{t}}}{T \cdot P E_{\mathrm{total}}}=\frac{Y_{CONV}}{T \cdot P_{SA1}  P_{SA2}}
\end{equation}
\end{footnotesize}
where $T$, $Y_{CONV}$ denote the total consumed cycles and the total number of required multiply-accumulate operations for one CONV layer, respectively.  
$PE_{\mathrm{on}_{t}}$ is the number of effective working PEs in cycle $t$ that contributes to effective $Y_{CONV}$ computations. $PE_{\mathrm{total}}$ is the total number of PEs.

Figure \ref{fig:opt1_i4},\ref{fig:opt1_gn} show the values of $\mu_i$ for each CONV layer $i$ in Inception-v4 and GoogleNet. In each figure, 
we use the algorithm mapping returned by the framework.
The "square-NS" ($bl_1$) plot shows the theoretical layer-wise effective PE utilization assuming the algorithms are deployed with the stall-free PE and data layout optimizations on the \textit{largest square-shaped systolic array} within the resource constraint on the target FPGA, but no dataflow optimization is applied and only NS dataflow is used across all layers. Considering the upper DSP bound - 6084=78$\times$78 - input into the framework, such largest systolic array is shaped at (78,78). The "algo1-NS" plot ($bl_2$) evaluates the effect of systolic-array dimension identification as performed using Algorithm \ref{algo: hardware_algo}, without dataflow optimization.
The "algo1-optimized" ($OPT$) plot shows the resulting effective PE utilization when running the same set of algorithms using the ($P_{SA1},P_{SA2},\psi$) identified by DYNAMAP (Algorithm \ref{algo: hardware_algo}). Compared to $bl_2$ which only uses NS dataflow, $OPT$ shows consistent improvement on almost all the layers as it minimizes zero-paddings required in the systolic array as discussed in Section \ref{sec:dfpe}. Compared to $bl_1$ which uses a square-shaped systolic array, although for a very small number of layers the $OPT$ utilization is not as saturated, the performance loss in those layers are much smaller than the performance gains obtained in other layers.
 Algorithm \ref{algo: hardware_algo} finds the sweet spot for the shape of the systolic array that drives up $\mu$ across all the layers and minimizes end-to-end latency.
 Global CNN wide trade-off analysis performed by $OPT$ ensures that the performance benefit on more compute-intensive layers overweights any losses on other layers. Overall, compared to a NS-dataflow implementation on the largest square-shaped systolic array, by implementing the designs generated by DYNAMAP, we observe $32\%$ and $35\%$ lower latency in end-to-end latency for Googlenet and Inception-v4, respectively.

\vspace{-2mm}
\subsubsection{Effect of layer-wise algorithm switching}
\label{algoexpsec}
\begin{figure}[ht]
    \centering
    \includegraphics[width=8cm]{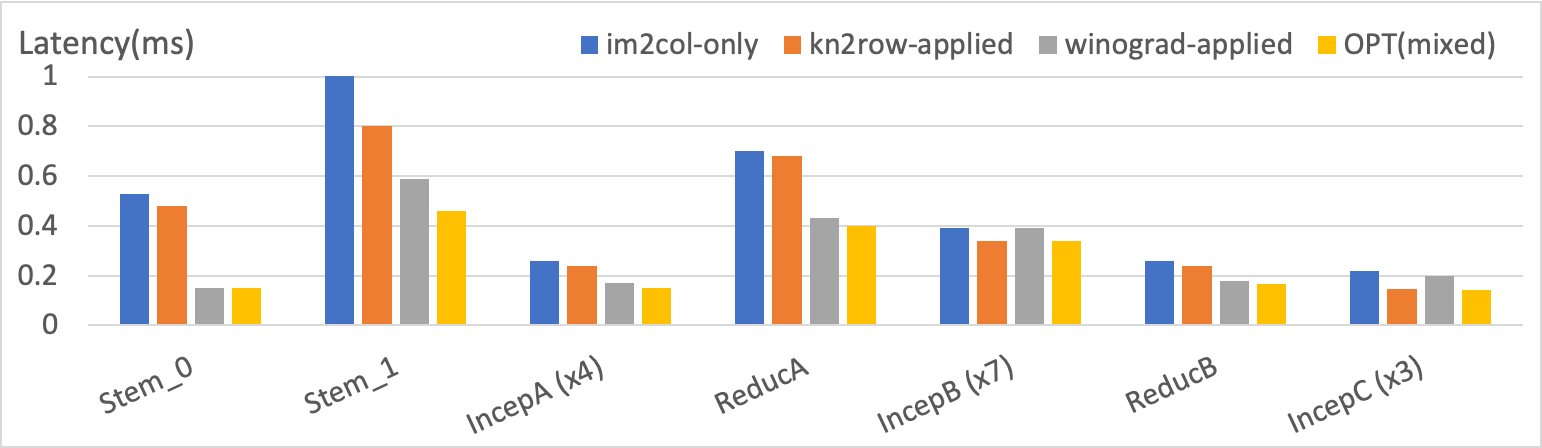}
    \vspace{-3mm}
    \caption{Layer exe. times: Inception-v4}
    \label{fig:opt3_i4}
\end{figure}

\begin{figure}[ht]
    \centering
    \includegraphics[width=8cm]{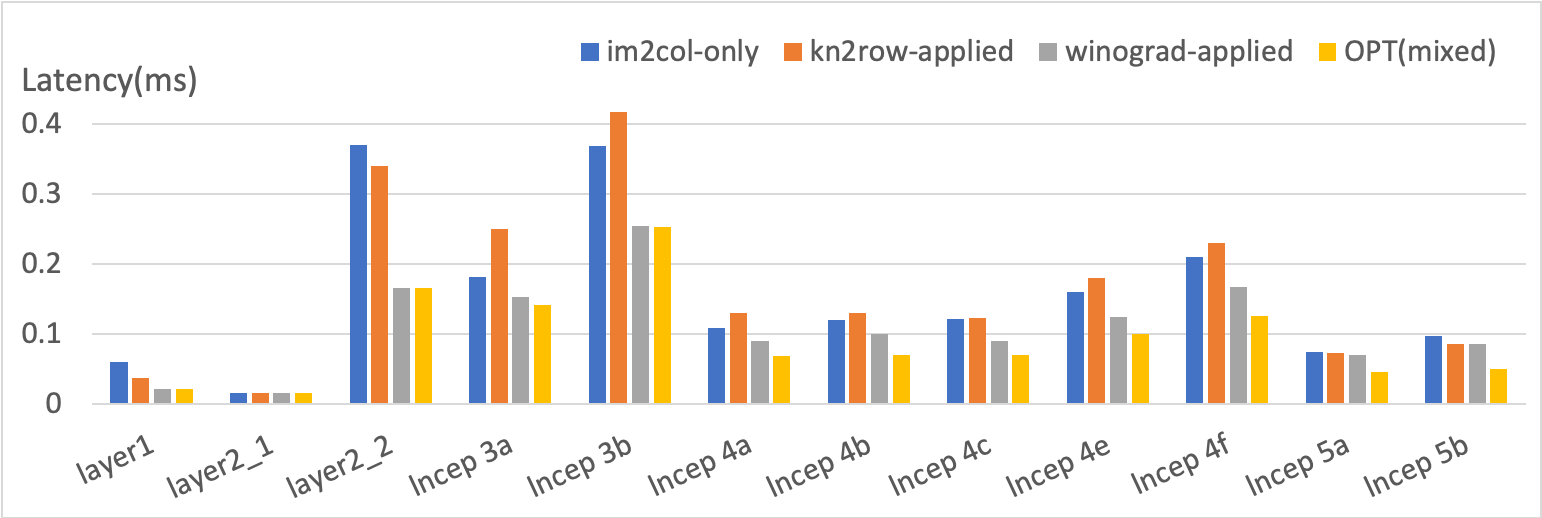}
    \vspace{-3mm}
    \caption{Layer exe. times: GoogleNet}
    \label{fig:opt3_gn}
\end{figure}

We calculate the execution time of each CNN module using different algorithms. Figure \ref{fig:opt3_i4} and \ref{fig:opt3_gn} shows the results for the two CNNs, where on the x axis we group all the CONV layers in each Inception (Reduction) Modules, consistent with the notions in \cite{szegedy2015going, szegedy2017inception} and the corresponding columns show the sum of computation and communication latency of all layers in an Inception (Reduction) Module. The STEM module in Inception-v4 is broken down by the first Filter Concatenation layer for better visibility. The "im2col-only" ($bl_3$) columns show the result of using one algorithm - im2col - across all the layers, the "kn2row-applied" ($bl_4$) columns show the results of applying kn2row where possible and im2col everywhere else, and the "wino-applied" ($bl_5$) columns show the results of applying Winograd ($m=2,r=3$) where applicable (i.e. layers with square-shaped kernels) and im2col everywhere else. $OPT_{returned}$ are the results
using algorithm mapping returned by DYNAMAP, which is observed to be superior than all $bl_{3-5}$ on all modules. In Inception-v4, a large portion of the kernels are shaped $7(3)\times 1$, 
making such layers more memory-bound, therefore 
kn2row almost always out-perform im2col, which requires data duplication and results in less data-reuse. However, on GoogleNet, for most layers the lower-communication-cost benefit of kn2row do not offset the overheads due to "Pad-and-Accumulate" and serializing a large GEMM into $K^2$ smaller ones, making it less advantageous. A typical GoogleNet Inception Module has two layers with square-shaped $3\times 3$ and $5\times 5$ kernels among others. While applying Winograd on such layers always reduces the computation complexity, it is not always optimal overall. This is because for kernels larger than $3\times 3$, $\frac{K_1K_2}{3^2}$ rounds of Winograd is required, resulting in severe transformation overheads and amortized decrease in computation complexity reduction. Winograd also imposes high memory overheads and layout transformation cost, so kn2row is overall better for such layers with slightly higher computation cost but significantly lower communication cost. These observations suggest that an algorithm mapping scheme that greedily chooses the algorithm with the smallest layer node cost $c$ would not return the optimal mapping. DYNAMAP captures the tradeoffs that occur in such algorithm transitions, yielding lower end-to-end latency than using any of the algorithm or even all three algorithms greedily selected based on layer node costs. 
The algorithm mapping obtained in DYNAMAP is optimal on the given systolic array (as supported by Theorem \ref{thm1}) and is obtained within 2 seconds on an AMD 3700X cpu.
The overall percentage decrease in the latency of the designs returned by DYNAMAP 
compared to the base-lines ($bl_{3-5}$) are summarized in Table \ref{algexpdyn}.
\begin{footnotesize}
\begin{table}[ht]
\caption{End-to-end Latency Improvement due to Dynamic Algorithm Mapping}
\vspace{-3mm}
\begin{tabular}{cccc}
\hline
\textbf{}             & $bl_3$ & $bl_4$ & $bl_5$ \\ \toprule
\textbf{GoogleNet}    & 67.5\%               & 78\%                    & 22\%                      \\ \midrule
\textbf{Inception-V4} & 86\%                 & 61\%                    & 17\%                      \\ \bottomrule
\end{tabular}
\label{algexpdyn}
\end{table}
\end{footnotesize}
\vspace{-2mm}
\subsection{Comparison with State-of-the-art}
\label{sota}
Table \ref{tab: exp comparison} compares the performance of our design produced by DYNAMAP with the state-of-the-art. We achieve 286MHz frequency for both GoogleNet and Inception-v4 accelerator designs. GoogleNet acceleration using DYNAMAP significantly outperforms \cite{gn_0} and \cite{gn_1} in terms of both latency and throughput. This is partly due to the advantage of DYNAMAP's optimizations on dataflow and algorithm switching, partly due to the lower-precision we adopted enabling more PEs. Even if we scale down the systolic array size (2 DSP consumption per PE), 
in the worst case the performance will be halved and we still achieve $2\times$ and $1.4\times$ lower latency compared with \cite{gn_0} and \cite{gn_1} respectively. For Inception-v4, we compare with \cite{incepv4} which applies dynamic memory management to overcome data transfer bottlenecks and \cite{incep_v4_2} that uses kn2row method for all layers in GoogleNet. Compared to \cite{incep_v4_2}, even with lower frequency, our design achieves 20\% speedup. While using Winograd on some layers leads to low complexity, its impact is limited as there are more memory-bound than computation-bound layers in Inception-v4. However, DYNAMAP allocates kn2row to those memory-bounded kernels while keeping the computation-bounded layers optimized as well, integrating dataflow optimization to improve hardware utilization in both cases. As CNNs evolve to be more layer-diverse and the tradeoffs become less obvious, the benefits of using DYNAMAP will become much more pronounced.

The motivation of FlexCNN~\cite{sohrabizadeh2020end} is similar to that of DYNAMAP. However, it uses dynamic tiling with data layout optimizations across different layers to drive up effective DSP utilization to as high as 93.5\%/91.4\% on 3x3-/1x1-kernel layers on the Open Pose-v2 network (2.9 GOPS). It achieves a single-image inference latency of 24.7ms using 8x8x8 systolic array. To estimate the best-case performance using FlexCNN to accelerate Googlenet ($\sim 3$ GOPS) and Inception-v4 ($\sim 9$ GOPS), we project this latency onto GoogleNet (Inception-v4) with 92x66(95x64) PEs as deployed in our design (optimistically assuming 100\% DSP utilization on all types of layers) :
$L_{projected-GN}=24.7ms\times \frac{8\times8\times8\times 93\%}{92\times 66\times 100\%}\times \frac{3 GOPS}{2.9 GOPS}=2 ms$,
$L_{projected-Incp4}=24.7ms\times \frac{8\times8\times8\times 93\%}{95\times 64\times 100\%}\times \frac{9 GOPS}{2.9 GOPS}=6 ms$, both higher than DYNAMAP's achieved latency. This is because DYNAMAP uses compute-reducing algorithm, Winograd, and memory-saving algorithm, kn2row, to resolve bottlenecks in 
both compute-intensive and memory-bound layers. The achieved performance benefits offsets the additional overheads for switching between different algorithms in DYNAMAP.

%% file: 8_conclusion.tex
\vspace{-2mm}
\section{Conclusion}
In this paper, we proposed an architecture-algorithm co-optimization framework to achieve low-latency CNN inference on FPGA. Our proposed hardware overlay includes several optimizations to achieve no-overhead dataflow switching and low-overhead algorithm switching. Our software tool flow achieves fast algorithm mapping and hardware customization. DYNAMAP has a wide applicability in optimizing the acceleration of any complex CNN models, even with extremely diverse layer configurations, on any FPGA devices. In the future, we will explore the possibility of generalizing DYNAMAP to a wider range of algorithms, including strided-Winograd and frequency-domain methods.

\section{Acknowledgements}
This work was supported by US NSF under grant No. CNS-2009057 and No. OAC-1911229.